\documentclass{iau}
\usepackage{graphicx}
\usepackage{amsmath}
\usepackage{graphicx}
\usepackage{multirow}
\title[Piggybacking astronomical hazard investigations on Big Data science missions] 
{Piggybacking astronomical hazard investigations on scientific Big Data missions}

\author[Verdoes Kleijn et al]   
{Gijs A. Verdoes Kleijn$^{1,2}$,
Teymoor Saifollahi$^1$,
Rees Williams$^3$
Oscar Stolk$^1$,
Georg Feulner$^4$,
 }

\affiliation{$^1$Kapteyn Astronomical Institute, University of Groningen, The Netherlands \\ $^2$Netherlands Research School for Astronomy, The Netherlands, $^3$Donald Smits Centre for Information Technology, University of Groningen, The Netherlands, $^4$  Potsdam Institute for Climate Impact Research, Germany, 
}

\pubyear{2023}
\volume{274}  
\pagerange{xxx--yyy}
\jname{Astronomical Hazards for Life on Earth}
\editors{G. Tancredi}
\begin{document}

\maketitle

\begin{abstract}
Current and upcoming large optical and near-infrared astronomical surveys have fundamental science as their primary drivers. To cater to those, these missions scan large fractions of the entire sky at multiple wavelengths and epochs. These aspects make these data sets also valuable for investigations into astronomical hazards for life on Earth. 
The Netherlands Research School for Astronomy (NOVA) is a partner in several optical / near-infrared surveys. In this paper we focus on the astronomical hazard value for two sets of those: the surveys with the OmegaCAM wide-field imager at the VST and with the Euclid Mission.
For each of them we provide a brief overview of the astronomical survey hardware, the data and the information systems. We present first results related to the astronomical hazard investigations. We evaluate to what extent the existing functionality of the information systems covers the needs for the astronomical hazard investigations.
\end{abstract}

\firstsection 
\section{Introduction}

\begin{figure}[h]
   \centering
   \includegraphics[width=13cm]{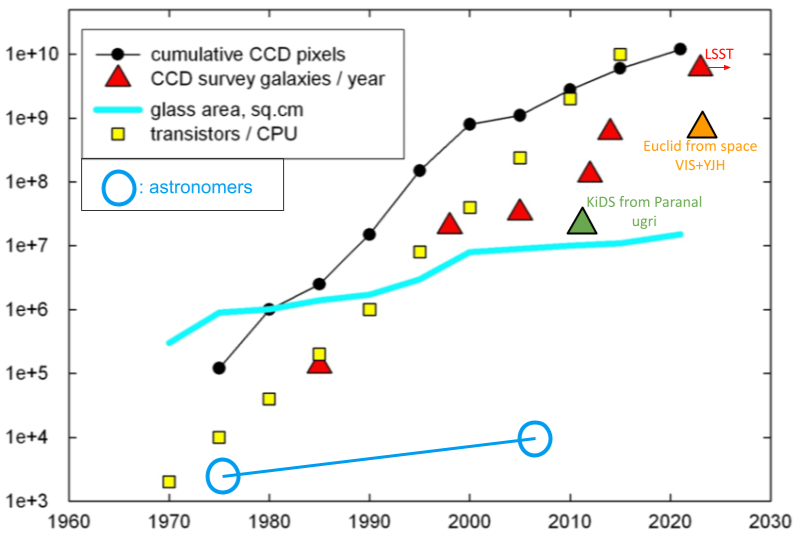}
   \caption{Growth of information technology, galaxy surveys and astronomers as a function of time. The typical number of transistors per CPU, the aperture area of telescopes ("glass"), the cumulative number of pixels in telescope cameras and number of astronomers are shown. The plot is taken from \cite{verdoeskleijn23} which adapted it from \cite{tyson19}. The triangles show for surveys the total number of galaxies observed with S/N $>$ 5 for flux inside a 2 arcsec wide aperture per year in at least one survey filter. The arrow for the LSST triangle indicates it is currently planned to start late 2024.}
   \label{HaystackGrowth}
\end{figure}

In the last four decades there has been an exponential growth in the observational data gathered by optical and near-infrared astronomical imaging surveys (see e.g., Fig.\,\ref{HaystackGrowth}, \cite{tyson19}, \cite{verdoeskleijn23}). This growth is continueing unabated. 
The Netherlands Research School for Astronomy (NOVA\footnote{https://nova-astronomy.nl/} ) is partner in several optical / near-infrared surveys. In this paper we focus on two sets of those: the surveys with the OmegaCAM wide-field imager at the VST, in particular the Kilo-Degree Survey and the ground-based and space-based surveys part of the Euclid Mission. These missions have fundamental astronomical science as their primary driver. For this they perform observations of large fractions of the entire sky at multiple wavelengths and at multiple epochs. These aspects make these data sets also valuable for investigations into astronomical hazards for life on Earth.
These survey missions will reach the tens of Terabytes regime in terms of catalogs and metadata databases and up to tens of Petabyte regime in terms of bulk data volume. This volume is spread over up to hundreds of thousands of exposures with each of them producing rich sets of metadata, such as catalogs. These in turn lead to millions of bulk data files. Therefore, the scientific exploitation and mining of these "Big Data" sets requires information systems which have an advanced databasing system at their core. Furthermore, to support the calibration of the raw data and its subsequent scientific analysis these systems also need to interface to high performance compute clusters and massive storage systems. 
For both survey missions NOVA has a leading role in the development and operation of the associated information systems. This offers NOVA a good opportunity to (re-)use both the observational data and the associated information systems for investigations into astronomical hazards, such as posed by asteroids, comets and close stellar encounters. This way NOVA can make a contribution to protecting society against astronomical hazards, on short and long terms. 
In this paper we describe two recently initiated pilot projects within NOVA investigating astronomical hazards that piggyback on the survey data and associated information systems developed for fundamental science. One project pertains to Near-Earth Objects using OmegaCAM data with the \textsc{AstroWISE} information system. The other project pertains to climate change due to comet impacts due to close stellar encounters. This will use the surveys of the Euclid Mission and its Euclid Data Processing System. For both each we present first results related to the astronomical hazards. We also provide an overview of the astronomical survey hardware, the data and the information systems. In this we highlight to what extent the existing functionality of the information systems covers the needs for the astronomical hazard investigations.

\section{Near-Earth Object precovery and discovery with OmegaCAM and AstroWISE}

Near-Earth Objects (NEOs) are asteroids or comets whose perihelion occurs at less than 1.3 astronomical units (au), meaning that close approaches with the Earth might occur at some point. The size of these objects ranges from meters to tens of kilometers. The impact hazard poses a significant threat to life on Earth (\cite{Borovicka13}, \cite{Popova13}, \cite{Brown13} ) and calls for planetary defence strategies. The Tunguska and Chelyabinsk impacts are recent reminders. Characterization of the orbits and physical composition of Near-Earth Objects is thus valuable for planetary defence. Serendipitous recovery of NEO appearances in archival scientific astronomical observations not dedicated to Near-Earth Objects can contribute to this characterization. 
In particular, NEO \textit{precovery} $-$ detection of a known NEO in an observing dataset prior to its discovery $-$ provides kinematic information about the NEO at a location in its orbit possibly valuable in addition to the discovery and immediate follow-up observations. This is because the orbit uncertainty depends on the fraction of the orbital arc that is covered during discovery and follow-up. Precovering one or more points far away from the discovery and immediate follow-up observations can significantly improve the accuracy of the orbital parameters. In this way, science-driven missions not only have a scientific purpose but can also provide a societal spin-off. 

Therefore we initiated with ESA and within NOVA an exploratory pilot that evaluates both the re-use of astronomical imaging surveys, in this case those of the OmegaCAM wide-field imager, and the re-use of the information system which was developed to handle the production and scientific analysis of these surveys: \textsc{AstroWISE}. 

\begin{figure}[h]
    \centering
    \includegraphics[width=\linewidth]{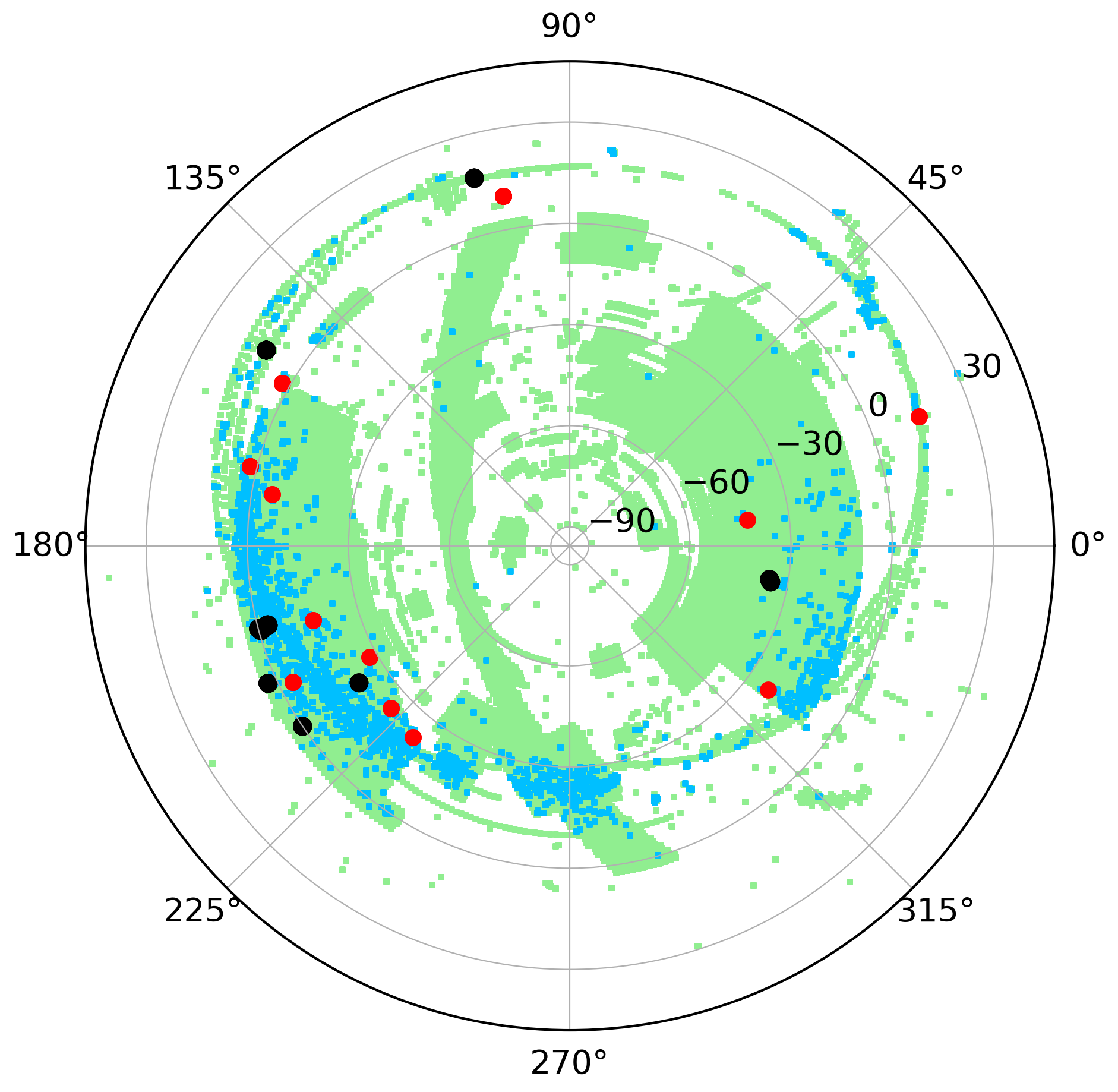}
    \caption{The sky coverage of the over 400 000 OmegaCAM/VST observations (green area). Of those about 10 000 frames are predicted to overlap with a known NEO (light blue dots). For the small subset predicted to have sufficient signal-to-noise, those with a successful and failed detection are shown as black and red points, respectively.}
    \label{ocam-coverage-2}
\end{figure}

OmegaCAM is a wide-field camera on the VLT Survey Telescope at ESO's Cerro Paranal Observatory. OmegaCAM has 32 science CCDs with a field of view of approximately 1 square degree. Over the first decade of its operation, OmegaCAM has covered a significant portion of the southern hemisphere (Fig~\ref{ocam-coverage-2}) in over 400 000 exposures.

\textsc{AstroWISE} stands for Astronomical Wide-field Imaging System for Europe which is an information system for data management, image processing, and calibration for a range of astronomical telescopes and instruments in a single data flow environment (\cite{aw2}, \cite{awe}). It has been used to do survey production for example for OmegaCAM's Kilo-Degree Survey (\cite{kuijken19}) and the Fornax Deep Survey (\cite{peletier20}). 

For the pilot we developed an AstroWISE Precovery Pipeline which re-used the \textsc{AstroWISE} functionality for data calibration, processing and analysis and added automated interfaces to webservices for ephemerides prediction (SSOIS, \cite{Gwyn}, JPL Horizons\footnote{https://ssd.jpl.nasa.gov/horizons/app.html}) and includes the deployment of dedicated software for the detection of streaks (\textsc{StreakDet}, \cite{streakdet}, \cite{streakdet2}).

The pilot with the AstroWISE Precovery Pipeline resulted in the recovery of 196 appearances of NEOs from a set of 968 appearances predicted to be recoverable. The achieved astrometric and photometric accuracy is on average 0.12arcsec and 0.1\,mag. It includes 49 appearances from a set of 68 NEOs predicted to be recoverable and which were on ESA's and NASA's risk list at that point.
ESA's risklist\footnote{https://neo.ssa.esa.int/risk-list} is provided by the ESA near-Earth Objects Coordination Centre (ESA-NEOCC) and consists of known NEOs with a non-negligible chance of impact in the next hundred years.
The appearances of three NEOs constituted precoveries, i.e., appearances well before their discovery. The subsequent risk assessment using the extracted astrometry removed these NEOs from the ESA and NASA risk list. For an in-depth discussion of the methods and results we refer to \cite{saifollahi23}. 

Using the experience of the pilot we attempt here to answer questions on the value and challenges of re-using the astronomical data and associated information systems for planetary defence against NEOs.

\textbf{What is the detectability of NEO appearances in astronomical archives such as the OmegaCAM archive?} 
We define NEO detectability as the fraction of detectable appearances among the total of occurrences that a NEO is predicted to be located within the FoV of the images. For the OmegaCAM pilot, the detectability varies as a function of the chosen threshold of signal-to-noise (SNR). The detectability rate is estimated to be $\sim$0.005 at an SNR$>3$ for NEOs on the risk-list and for the full list of NEOs. We expect no significant improvement can be made in detectability given the low 3$\sigma$ threshold in predicted SNR and the fact that detected NEOs tend to be often a few tenths of magnitude fainter than predicted for this pilot. \\

\textbf{What is the precovery rate for NEOs predicted to be detectable?}
The precovery rate for SNR$>$3 is 40\% for NEOs on the risk-list and 20\% for the full list of NEOs. The precovery rate increases to about 50\% for SNR$>$10. So a factor of up to 5 more NEOs can be precovered from the OmegaCAM archive through improved detection techniques (see below for discussion on new techniques). It will be a significant result as well if, after improving the recovery processes,  the failed recoveries turn out to be in fact non-appearances. It would suggest that the actual orbital accuracy for those objects (including those on the risk list) is significantly worse than predicted. \\
    
\textbf{What astrometric and photometric accuracy can be achieved?} 
The astrometric and photometric accuracies are 0.12arcsec (15\% of the average FWHM of OmegaCAM/VST images of about 0.8arcsec) and 0.1\,mag. Improvements in astrometric accuracy are expected from propagating the proper motions in the Gaia astrometric reference catalog to the observation date of the science image. Improvements in photometric accuracy can come from a more sophisticated modelling of SED, observational configuration and NEO shape modelling. The Solar System Open Database Network might facilitate this  (\cite{berthier22}).\\
    
\textbf{What are the challenges in deploying the Precovery Pipeline, also on imaging archives from other instruments available in \textsc{AstroWISE}?}
The Pipeline works mostly automatically through all steps to produce candidate recoveries. These are then inspected by an expert for confirmation/rejection. Thanks to the common data model for calibrated observations in \textsc{AstroWISE} (\cite{mcfarland13}) it could be deployed straightforwardly on calibrated observations for the imaging archives of about a dozen other cameras available in \textsc{AstroWISE}.  A challenge is that precise photometric calibration for a range of instruments is hard to fully automate. This is because the derivation of the solution requires reference stars sometimes inside the science images, sometimes in separate calibration observations. A potential solution would be to construct a photometric reference catalog that spans the entire sky observable by OmegaCAM with sufficient stellar density. This appears possible by aggregating information from the multiple large-scale surveys of the Southern Sky. Another main challenge is robust NEO detection and segmentation. This is also a main reason behind the the obtained recovery rates. \textsc{StreakDet} is a great tool for detecting high SNR streaks with sizes between 5-20arcsec. However, its performance drops for faint and long streaks. Deep learning might be a solution to improve streak detection and ultimately NEO precovery (e.g., \cite{pontinen2020}). \\

For an in-depth discussion about the recovery results for OmegaCAM and the feasibility of deploying it to other instruments, we refer the reader to \cite{saifollahi23}.

\section{Close stellar encounters with Euclid surveys and systems}

\begin{figure}[!htb]
    \centering
    \includegraphics[width=\linewidth]{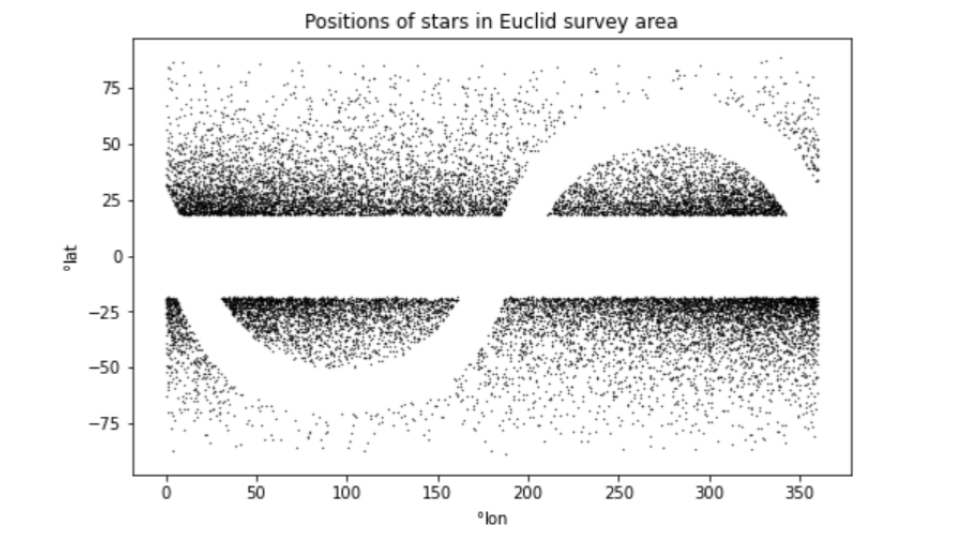}
    \caption{The position of stars in a simulated Milky Way distribution in Galactic coordinates. An Euclid's primary science driver is galaxy weak lensing tomography. Therefore its survey area is at galactic latitudes larger than |23 deg| to avoid the Milky Way disk and at ecliptic latitudes larger than |10 deg| to avoid near-infrared background contamination by zodiacal light. The simulation is made using the Galaxia modeling code (\cite{sharma11}) in its wrapper code (\cite{rybizki18})}
    \label{EuclidStarsSimulation}
\end{figure} 

\begin{figure}[!htb]
    \centering
    \includegraphics[width=\linewidth]{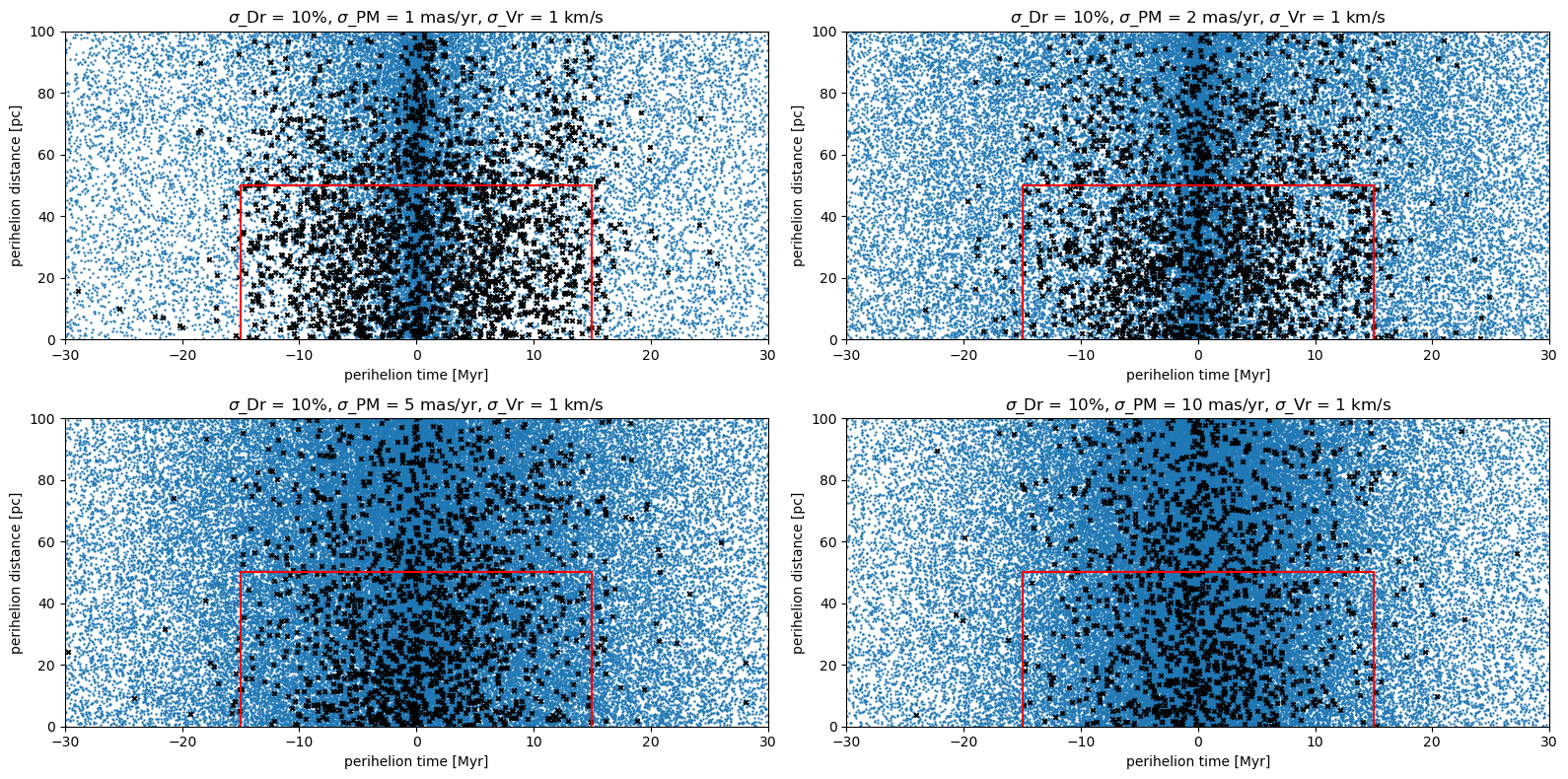}
    \caption{Estimated perihelion distances versus perihelion time for the stars in a simulated Milky Way distribution that can be detected by Euclid. This preliminary simulation selected randomly just 1\% of the simulated Milky Way stars before performing the linear motion approximation to estimate perihelion time and distance. The small blue dots show ten perihelion estimates for each simulated star by sampling ten times independent Gaussian random errors in distance ($\sigma$\_Dr), 2D proper motion ($\sigma$\_PM) and radial velocity ($\sigma$\_Vr). The large black dots show the true perihelion distance and time for stars that have an estimated perihelion to lie inside the red box of encounters within 50 pc within 15 million years. Completeness and false positive rate in a sample of candidate close encounters clearly depend critically on the measurement errors.}
    \label{StellarPerihelionPrediction}
\end{figure} 

Close encounters of stars to the Sun can affect climate and life on Earth. The ionizing radiation and cosmic rays from supernovae could have a significant impact on both for encounters within 10 pc (Thomas, these proceedings). Stellar encounters within 1 pc can cause significant gravitational perturbations in our Solar System's Oort Cloud. These can lead to increased influx of comets and hence planetary impacts in the inner Solar System  (Bailer-Jones, these proceedings). Close stellar encounters can also bring an increase in the influx of exocomets. They can originate in either the Oort cloud of the passing star or in the cloud's tidal streams  (\cite{portegieszwart21}). Impacts by comets and asteroids may have caused climate changes in the past (\cite{brugger17}) and might do so again in the future. 

Identifying close stellar encounters requires six dimensional phase space coordinates (three positions, three velocities) for stars in the Milky Way. The Gaia Mission has brought an enormous information leap in stellar phase space measurements. Its third data release provides an all sky astrometric reference frame sampled with almost 1.5 billion point sources down to 22nd magnitude (\cite{gaiadr322}). It also provides six-dimensional phase space coordinates for over 33 million stars down to G=14 with positions accurate at the milli-arcsecond (mas) level, proper motions at the mas / year level and radial velocities at the km/s level. This stellar sample allowed identification of 42 stars with encounters within 1 pc with a perihelion time up to ~6 Myr in the past and future (\cite{bailer-jones22}). From a similar analysis on Gaia's second data release it was estimated that about 15\% of all close stellar encounters within 5 pc and within 6 Myr were detected at that point (\cite{bailer-jones18}).  The associated inferred rate of encounters within 1 pc is about 20 per million year. The final Gaia release might roughly double the completeness and  be able to detect encounters with perihelion times of order 10 Myr in past and future. 

To increase this completeness level and perihelion time span one has to identify close encounters from stars fainter than observable by Gaia using deeper surveys. Five dimensions of the six-dimensional phase space can be obtained by combining imaging surveys observing the same sky area at multiple wavelengths and at multiple epochs. The multiple epochs allow to derive proper motions (in addition to the positions). The multiple wavelengths allow to derive photometric distances (see e.g., Chapter 3 in \cite{speagle20}). ESA's Euclid Mission brings together space-based and ground-based surveys at multiple epochs and multiple wavelengths over almost 15 000 square degrees of sky. Observations at 9 wavelengths are gathered via 8 instruments, located at 7 telescopes in space and on the ground. The first observations which are now being re-used as part of the Euclid Mission occurred from the ground in August 2013. The space-based observations for the Euclid Mission will be obtained with ESA's Euclid satellite. It will be launched in July 2023 and observe the almost 15 000 square degrees of extragalactic sky in about 6 years (\cite{laureijs11}, \cite{scaramella}). The ground-based observations are planned to be completed well before July 2029. 
The Euclid satellite will survey the extragalactic sky using a 1.2m telescope with two imagers. The visible imager (VIS) and Near Infrared Spectrometer and Photometer (NISP), sharing a 0.53 square degree Field of View. VIS will detect point sources down to a limiting magnitude of 25 (AB, 10$\sigma$ for a point source measured using a 2 arcsec diameter aperture) using a very broad filter (550–900 nm). The Near-infrared Spectrograph and Photometer (NISP) will measure their photometry through Y, J, and H filters down to a magnitude limit of 23.5 (using same definition as VIS). All space observations of a sky area will be done at a single epoch. This data will be combined with data from ground-based telescopes in the optical filter u, g, r, i, z to matching depth.  In the Northern hemisphere this will be with four surveys. The Canada-France Imaging Survey (CFIS, \cite{ibata17}) observes in u and r. CFIS observations started in the first semester of 2015 and are done to full depth in a single epoch. The Waterloo Hawaii IfA G-band Survey (WHIGS\footnote{https://www.skysurvey.cc/aboutus/}) observes in g. WHIGS started approximately 2022 and observes to full depth in a single epoch. The Panoramic Survey Telescope And Rapid Response Systems 1 and 2 (Pan-STARRS 1 \& 2, \cite{kaiser10}) observes in i band. Pan-STARRS observations cover the Euclid survey area since 2010, building up depth by many revisits over years until a few years after 2023. The Wide Imaging with Subaru HSC of the Euclid Sky (WHISHES\footnote{https://www.skysurvey.cc/aboutus/}) survey observes in z. It observes since the second semester of 2020 to full depth in single epochs. In the Southern hemisphere the Euclid survey area is covered by the Dark Energy survey in g, r, i and z (DES, \cite{abbott21}) and the Large Survey of Space and Time (LSST, \cite{ivezic19}) in u, g, r, i and z. DES observed from August 2013 until January 2019, building up depth in yearly revisits. The Vera Rubin Observatory plans to start the LSST survey late 2024 and has 10 years of planned operations. It will build up depth through many visits over many years. 

Combining such a heterogeneous set of epochs and filters into a homogeneously set of order billion stellar positions, proper motions and distances requires a careful calibration approach using a information system that also allows ample quality control. The astrometric calibration might best be done via calibration against with zero proper motion and well-defined and consistent centroids across the optical and near-IR. For this reason \cite{tian17} used compact galaxies as calibrators. They combined Gaia (Data Release 1) with data from the SDSS, 2MASS and PanSTARRS surveys to obtain proper motions for 350 million sources with a characteristic systematic error of less than 0.3 mas/year and a typical precision of 1.5–2.0 mas/year. The Euclid survey area will contain of order a billion stars. For Euclid such an astrometric calibration effort can be performed and released multiple times as observations on space and ground progress. To do such a massive operation repeatedly that accurately for so many objects is facilitated by an advanced databasing system providing a rich and detailed description of all data items (\cite{mulder20}). The Euclid information system (called the Euclid Archive System) is such a "data-centric" system. It consists of two main components: the Euclid Science Archive System and the Euclid Data Processing System (\cite{nieto19}). All bulk and metadata of calibration and science observations required for the re-use to determine close stellar encounters reside in the Data Processing System. Only a subset resides in the Science Archive System. For example, all bulk data, metadata and data quality reports related to individual ground-based exposures resides only in the Data Processing System. Combining Euclid's five-dimensional phase space (positions, distances and proper motions) with stellar radial velocities, from e.g., a spectral survey, establishes then finally the six-dimensional data set from which one can infer which of these billion stars (mostly fainter than observable by Gaia) lead/led to close stellar encounters. 


Simulations are on going to determine what completeness and accuracy to expect in terms of nearby stellar encounters using Euclid Mission's observations gathered over almost two decades at 9 wavelengths when combined with such a radial velocity survey. Fig~\ref{EuclidStarsSimulation} shows a simulated Milky Way stellar distribution as observed by Euclid. The simulation is made using the Galaxia modeling code (\cite{sharma11}) in its wrapper code (\cite{rybizki18}). The code simulates magnitudes in all 9 Euclid filters and 6D phase coordinates. 
Photometric distance estimates can be derived using the magnitudes as input to a Bayesian statistical framework and
modeling (e.g., \cite{speagle20}). Proper motions could be obtained from the positional catalogs of all observations following the approach of \cite{tian17}. Finally stellar radial velocities have to be supplied by another mission than Euclid. The perihelion distance and time of close stellar encounters can then be estimated using the Linear Motion Approximation (\cite{bailer-jones18}). Very preliminary estimates are shown in Fig~\ref{StellarPerihelionPrediction} for four different assumptions on accuracy of estimated distances, proper motions and radial velocities. Completeness and false positive rate in a sample of candidate close encounters clearly depend critically on the measurement errors.



\section{Lessons learned}

Above's two pilots have shown to us that it is worthwhile to explore further the piggybacking of astronomical hazard investigations onto the data and information systems developed for astronomical scientific Big Data missions. During the execution of above's two pilots we noticed two general characteristics of the astronomical scientific information systems that are key in making them more amenable for re-use in astronomical hazard investigations: 

\begin{itemize}
    \item Standardization is important. Using common interfaces, data models and data structures inside a single information system / environment for the datasets from different instruments makes it significantly less effort to develop and implement pipelines for astronomical hazard investigations.   
    \item "Data-centrism" is important. Survey systems come in two main flavors. There are data-centric ones: those that put a rich and detailed specification of the data items/objects at the core of their architecture and at the "fingertips" of the user of the system. Such systems are in a sense the system analog of object oriented programming. There are the process-centric ones: those that put a rich and detailed specification of the data processing at the core of their architecture and at the "fingertips" of the user of the system. They are the system analog of functional programming. For the re-use of data items/objects for other purposes (like astronomical hazards), the rich and detailed specification of data items at the fingertips of the user make data re-use convenient. In the case of processing-centric architectures re-use of data becomes cumbersome if the dominant re-use is not in the re-use of processes.   
\end{itemize}


\section{Acknowledgments}
This work was executed as part of ESA contract no. 4000134667/21/D/MRP (CARMEN) with their Planetary Defence Office. The pilots made use of the Big Data Layer of the Target Field Lab project "Mining Big Data". The Target Field Lab is supported by the Northern Netherlands Alliance (SNN) and is financially supported by the European Regional Development Fund. The data science software system \textsc{AstroWISE} runs on powerful databases and computing clusters at the Donald Smits Center of the University of Groningen and is supported, among other parties, by NOVA (the Dutch Research School for Astronomy).  This research has made use of Aladin sky atlas (\cite{aladin1,aladin2}) developed at CDS, Strasbourg Observatory, France and SAOImageDS9 (\cite{ds9}). This work has been done using the following software, packages and \textsc{python} libraries: Astro-WISE (\cite{aw2}, \cite{awe}), \textsc{Numpy} (\cite{numpy}), \textsc{Scipy} (\cite{scipy}), \textsc{Astropy} (\cite{astropy}).

\bibliography{iaus374_verdoeskleijn.bib}

\end{document}